\documentclass[aps,twocolumn,pra]{revtex4}

\usepackage{epsf}
\usepackage{amsfonts}
\usepackage{amssymb}
\usepackage{graphicx}
\usepackage{color}
\usepackage{amsmath}
\usepackage[bookmarksnumbered, bookmarks, breaklinks, linktocpage]{hyperref}

\newcommand{\bra}[1]    {\langle #1|}
\newcommand{\ket}[1]    {| #1 \rangle}

\newcommand{\kb}[2]     {| #1 \rangle \! \langle #2 |}

\newcommand{\cA}        {{\mathcal A}}
\newcommand{\cE}        {{\mathcal E}}

\newcommand\hocom[1]{}

\newcommand{\ba}{\begin{eqnarray}}
\newcommand{\ea}{\end{eqnarray}}
\newcommand{\bmath}{\begin{mathletters}}
\newcommand{\emath}{\end{mathletters}}
\newcommand{\ban}{\begin{eqnarray*}}
\newcommand{\ean}{\end{eqnarray*}}

\begin{document}

\title{Quantum origin of quantum jumps: Breaking of unitary symmetry induced by information transfer in the transition from quantum to classical }
\author{Wojciech Hubert Zurek}

 \address{Theory Division, MS B213, LANL
    Los Alamos, NM, 87545, U.S.A.}

\date{\today}

\begin{abstract}
Measurements transfer information about a system to the apparatus, and then further on -- to observers 
and (often inadvertently) to the environment. I show that even imperfect copying essential in such 
situations restricts possible unperturbed outcomes to an orthogonal subset of all possible states 
of the system, thus breaking the unitary symmetry of its Hilbert space implied by the quantum superposition principle. Preferred outcome states emerge as a result. They provide framework 
for the ``wavepacket collapse'', designating terminal points of quantum jumps, and defining the
measured observable by specifying its eigenstates. In quantum Darwinism, they are the progenitors of multiple copies spread throughout the environment -- the fittest quantum states that not only survive 
decoherence, but subvert it into carrying information about them -- into becoming a witness.

\end{abstract}

\maketitle

\section{Introduction: Quantum Axioms}
The quantum principle of superposition applies to isolated systems, but is famously violated 
in the course of measurements: A quantum system can exist in any superposition, but a measurement 
forces it to choose from a limited set of outcomes represented by an orthogonal set of states [1]. 
In textbook discussions of quantum theory these states are preassigned as ``the eigenstates of the measured observable''. They are the only possible states that can ever be detected (and therefore 
prepared) by that measuring device. As Dirac puts it [2]  ``...a measurement always causes 
the system to jump into an eigenstate of the dynamical variable that is being measured..." 

I show -- using ideas that parallel the no-cloning theorem -- that this restriction (usually imposed ``by decree'', by the {\em collapse postulate}) can be {\em derived} when a transfer of information essential for both measurement and decoherence is modeled as a unitary quantum process that leads to records with predictive significance. This sheds new light on quantum measurements, on the quantum - classical transition, and on the selection of preferred pointer states by the environment - induced decoherence [3-6]: It shows that a restriction to a limited set of orthogonal states (but, of course, not the non-unitary collapse {\em per se}) can be deduced in a setting that does not assume {\em a priori} existence of macroscopic apparatus usually invoked to define the measured observable [1,2,7]. This connection between the transfer of information and the selection of preferred states is also of crucial importance for quantum Darwinism [5], where the capacity of the environment to acquire multiple records of selected states of the system is essential. Resilient states that can withstand monitoring by the environment without getting disturbed become endowed with objective reality: They are simultaneously accessible to many observers through the imprints they leave in the environment.

Quantum theory is founded on several simple postulates [2,4-8]. The first two encapsulate the quantum principle of superposition and decree unitarity of evolutions. Thus; (i) {\it Quantum state of a system is represented by a vector in its Hilbert space} and; (ii) {\it Quantum evolutions are unitary (e.g., generated by the Schr\"odinger equation)} describe mathematical formalism of the theory. To make contact with the real world one needs to relate abstract unitarily evolving quantum states to experiments. The next postulate; (iii) {\it Immediate repetition of a measurement yields the same outcome}, is the first 
of the `measurement postulates'. It is uncontroversial -- states of classical systems satisfy it. It establishes predictive utility of quantum states that have already entered through postulate (i). One can rephrase predictability postulate by noting that when a state is known (for example, because of an earlier
measurement), then one can choose a measurement that will confirm it but leave it intact. 

These first three postulates indicate no bias -- they treat every state in the Hilbert space of the system 
on equal footing. In particular, postulates (i) and (ii) emphasize linearity: Any superposition of quantum states is a legal quantum state, and any evolution of such a superposition leads to a superposition 
of evolved ``ingredients''. By contrast, the last two postulates; (iv) {\it Measurement outcomes are restricted to an orthonormal set $\{\ket {s_k}\}$ of eigenstates of the measured observable} (i.e., measurement does not {\it reveal} the state of the system because it limits possible outcomes to the
preassigned outcome states), and (v) {\it The probability of finding a given outcome is $p_k=|\bra {s_k} \ket \psi|^2$, where $\ket \psi$ is the preexisting state of the system}, are at the heart of the long-standing disputes on the interpretation of quantum theory [1-10]. 

As a consequence of (iv) (the ``collapse postulate'') it is impossible to find out an unknown state of a quantum system: In contrast to the egalitarian postulates (i)-(iii), explicit symmetry breaking implied by (iv) defies the superposition principle by allowing only certain states as outcomes. 
The aim of this paper is to point out that already the (symmetric and uncontroversial) postulates (i)-(iii) necessarily imply selection of  {\it some} preferred set of {\it orthogonal} states -- that they impose the broken symmetry that is at the heart of the collapse postulate (iv) -- although they stop short of specifying what this set of outcome states is, and obviously cannot result in anything explicitly non-unitary 
(e.g., the actual ``collapse''). 

 \section{Predictability and symmetry breaking}

To see how symmetry - breaking restriction to a set of orthonormal outcome states arises from 
symmetric and uncontroversial assumptions (i)-(iii) consider a system ${\cal S}$, to be 
measured by an apparatus ${\cal A}$. For simplicity, we start with the smallest possible ${\cal S}$ 
with a two-dimensional Hilbert space ${\cal H}_{\cal S}$ (although our derivation will be extended to arbitrarily large systems). In a two dimensional ${\cal H}_{\cal S}$ any state of the system can be written as a superposition of a pair of linearly independent states, $\ket {\psi_{\cal S}} = \alpha \ket v + \beta \ket w $. This freedom to choose any two linearly independent $\ket v$ and $\ket w$ that need {\it not} be orthogonal is guaranteed by the principle of superposition implicit in postulate (i). A measurement of ${\cal S}$ by an apparatus ${\cal A}$ that starts in a ``ready to measure'' state $\ket {A_0}$:
\begin{eqnarray}
\ket {\psi_{\cal S}}\ket {A_0}=( \alpha \ket v + \beta \ket w )\ket {A_0} \Longrightarrow \nonumber \\
 \Longrightarrow \alpha \ket v \ket {A_v}+ \beta \ket w \ket {A_w} = \ket {\Phi_{\cal SA}} \ ,
\end{eqnarray}
yields the desired information transfer from ${\cal S}$ to ${\cal A}$: The state of ${\cal A}$ now contains a record of ${\cal S}$. 
Above we assumed postulate (iii) (there are states that are left intact, $ \ket v \ket {A_0} \rightarrow \ket v \ket {A_v}, \ \ket w \ket {A_0} \rightarrow \ket w \ket {A_w} $),
and  linearity that follows from postulate (ii). We have also recognized that the state of a composite quantum system (here ${\cal SA}$) is a vector in the tensor product of constituent Hilbert spaces (i.e, has a form of Eq. (1)), an observation often regarded as an additional ``complexity postulate (o)''. 

The norm of the state of the composite system must be preserved: Postulate (i) demands it (Hilbert space is a {\it normed} linear space), and postulate (ii) ensures that this demand is met. Thus,
simple algebra and the recognition that $\bra {A_0}\ket {A_0}=\bra {A_v}\ket {A_v}=\bra {A_w}\ket {A_w}=1$ yields
$$
\bra {\psi_{\cal S}}\ket {\psi_{\cal S}} - \bra {\Phi_{\cal SA}}\ket {\Phi_{\cal SA}} 
= 2 \Re \alpha^*\beta \bra v \ket w (1-\bra {A_v}\ket {A_w} )= 0 \ .
\eqno(2a) $$
Equation (2a) involves real part only, but must hold for arbitrary relative phase between $\alpha$ and $\beta$. This leads to:
$$ \bra v \ket w (1- \bra {A_v} \ket {A_w})=0 \eqno(2b)$$
This equality is the basis for our further discussion. 

Depending on the overlap $\bra v \ket w$ there are two possibilities. Let us first suppose that $\bra v \ket w \neq 0$ (but is otherwise arbitrary). In this case one is forced to conclude that the state of ${\cal A}$ cannot be affected by the process above. That is, the transfer of information from ${\cal S}$ to ${\cal A}$ must have failed completely, since $\bra {A_v} \ket {A_w}=1$ must hold: The apparatus can bear no imprint that distinguishes between the components of the superposition $\ket {\psi_{\cal S}}$ -- the prospective outcome states of the system.

The second possibility is that $\bra v \ket w = 0$. This allows for an arbitrary $\bra {A_v} \ket {A_w}$, including a perfect record, $\bra {A_v} \ket {A_w}=0$. Thus, outcome states $\ket v$ and $\ket w$ must be orthogonal if -- in accord with postulate (iii) -- they are to survive intact a successful information transfer in general or a quantum measurement in particular, so that the immediate remeasurement can yield the same result. The same derivation can be carried out for ${\cal S}$ with a Hilbert space of dimension ${\cal N}$ starting with a system state vector $\ket {\psi_{\cal S}}=\sum_{k=1}^{\cal N}\alpha_k \ket {s_k}$, where (as before), {\it a priori} $\{\ket {s_k} \}$ need not be orthogonal, but only linearly independent. 

Simple derivation above leads to very decisive conclusion: Orthogonality of outcome states of the system is absolutely essential for them to exert distinct influences -- to imprint even a minute difference -- on the state of any other system while retaining their identity: The overlap $\bra v  \ket w$ must be 0 {\it exactly} for $\bra {A_v} \ket {A_w}$ to differ from unity. Thus, also sloppy and accidental information transfers (e.g., to the environment during decoherence [3-6]) will  define preferred sets of states providing that the crucial predictability demand of postulate (iii) is met. This ``had to be so'':
When an imperfect measurement can be repeated many times without affecting
the original, collective records will come arbitrarily close to orthogonality.

Selection of an orthonormal basis induced by information transfer -- the need for a spontaneous symmetry breaking that arises from the otherwise symmetric axioms (i)-(iii) -- is a general and intriguing result. Our derivation parallels the proof of the no-cloning theorem [11-13]: We have employed the assumption of linearity and started with Eq. (1), as does the proof of no-cloning in Refs. [11] and [12]. Moreover, linearity and preservation of the norm follow from the unitarity postulate (ii) used in the alternative proof of no-cloning (see Ref. [13]; the only difference is that when copies are ``clones''  $\bra {v} \ket {w}=\bra {A_v} \ket {A_w}$ in Eq. (2b)). 

Similar reasoning appears in a proof of security of cryptographic protocols [14]. There, however, the focus is on the ability to detect eavesdroppers through perturbations her measurements inflict on the transmitted non-orthogonal states, which is rather different from the questions considered here. 
Nevertheless, connections between quantum prohibition on cloning, information gain and disturbance 
tradeoff (exemplified by Ref. [14]),  and our proof of orthogonality of outcome states are
hard to miss. However, implications we are led to are quite different.

The scope of our result is closer to the study of measurements due to Wigner [7]. He argued 
that record states of a {\it classical} apparatus must be orthogonal. He inferred this orthogonality from 
the ``Copenhagen assumption'' -- that an apparatus must be classical [1], and translated classical 
distinguishability of pointer positions into their orthogonality -- into $|\bra {A_v}\ket{A_w}| =0$. 
With the additional assumption of real eigenvalues, Wigner concluded that observables must 
correspond to self-adjoint operators. We did not use such a strong ``Copenhagen motivated" 
assumption -- $|\bra {A_v}\ket{A_w}| \neq 1 $ suffices to establish our result. Therefore, when we 
assume that eigenvalues should be real, Hermitean nature of observables immediately follows from 
axioms (i)-(iii), without any need to appeal to classicality of measuring devices.

The reader may be concerned about idealizations we have made to represent information transfer by 
Eq. (1). We now briefly explore these assumptions, see how they can be relaxed, and consider 
what this implies about terminal states for ``quantum jumps".

Let us start from the most obvious: The apparatus (or the environment ${\cal }$ monitoring the system 
in course of decoherence) is usually {\it not} in a pure state at the beginning, so one should represent 
it with a density matrix of the form $\rho^{\cal A}_0=\sum_k p_k \ket {a_k} \bra {a_k}$.  
To deal with this we note that such a mixed state can be always ``purified'' 
by allowing ${\cal A}$ to have a ``ghost partner'' ${\cal A}'$, so that the combined initial state of the two 
is given by $ \ket {{\bf A}_0} = \sum_k \alpha_k \ket {a_{0,k}} \ket {a_{0,k}'}$
in the obvious notation, and with $|\alpha_k|^2=p_k$. Orthogonality of the outcome states can be now established starting with the equation:
$$ \bra v \ket w(1- \bra {{\bf A}_v} \ket {{\bf A}_w})=0 \ . \eqno(2c)$$
As before, this leads to the conclusion that $\ket v$ and $\ket w$ (or any two outcome states that satisfy postulates (i)-(iii)) must be orthogonal if the information transfer is to succeed. Therefore, relaxing this assumption does not change our conclusions. This is also the case when the environment ${\cal E}$ is influenced directly by ${\cal S}$.

In particular, we did not invoke postulate (v) -- Born's rule -- which is necessary to attribute 
physical significance to reduced density matrices employed in the study of decoherence 
[3-6]. This is obvious when ${\cal A}$ is pure. Still, one might worry that ``purification" 
relating mixed state of ${\cal A}$ to pure state of enlarged ${\cal AA}'$ uses Born's rule 
``in reverse direction''. This is ``only mathematics'' ($p_k$ need not be regarded as probabilities), 
and it is not essential: It is easy to see that {\it any} prescription relating pure states of 
${\cal AA}'$ to mixed states of ${\cal A}$ will do, as long as it is based on one-to-one 
correspondence between eigenvalues of the density operator of ${\cal A}$ and absolute values of amplitudes in the entangled state of ${\cal AA}'$, and as long as it is used in both directions.
So even $|\alpha_k|^2=p_k$ (probability interpretation notwithstanding) is not needed.

This independence of unitary symmetry breaking responsible for quantum jumps from Born's rule 
is especially important when one aims to establish the extent to which ``controversial'' postulates 
(iv) and (v) follow from the generally accepted axioms (o)-(iii). In this context it is 
reassuring that the proof can be carried out without  ``purification''. To this end,
we note that unitary evolution preserves scalar product of operators defined as
$Tr \rho \tilde \rho$. Therefore, in the obvious notation, $Tr \ket v \bra v \rho_{{\cal A}_0} \ket w \bra w \rho_{{\cal A}_0} =
Tr \ket v \bra v \rho_{{\cal A}_v} \ket w \bra w \rho_{{\cal A}_w} $. Thus,
$$ |\bra v \ket w |^2 (Tr\rho_{{\cal A}_0}^2-Tr \rho_{{\cal A}_v} \rho_{{\cal A}_w})=0 \ . \eqno(2d)$$
So, either $\bra v \ket w = 0$ (we recover orthogonality of outcomes), or 
$Tr\rho_{{\cal A}_0}^2=Tr \rho_{{\cal A}_v} \rho_{{\cal A}_w}$. 
This last equality holds iff $\rho_{{\cal A}_v} = \rho_{{\cal A}_w}$. This is because 
$\rho_{{\cal A}_0} = \sum_k p_k \kb {a_{k0}} {a_{k0}}$, $\rho_{{\cal A}_v}=\sum_k p_k \kb {a_{kv}} {a_{kv}}$, and $ \rho_{{\cal A}_w}=\sum_k p_k \kb {a_{kw}} {a_{kw}}$ have the same eigenvalues $\{ p_k\}$
-- they are related to one another by a unitary evolution. Therefore, only their eigenvectors can differ. 
But then $Tr (\rho_{{\cal A}_0}^2-\rho_{{\cal A}_v} \rho_{{\cal A}_w})=\frac 1 2 Tr(\rho_{{\cal A}_v} - \rho_{{\cal A}_w})^2$, which is non-negative and vanishes iff $\rho_{{\cal A}_v} = \rho_{{\cal A}_w}$
-- only when there is no record left in $\cA$. 


This establishes that only orthogonal states can leave imprints in other systems without getting 
disrupted also in the case when the state of  that apparatus or the environment is mixed, and 
represented by a Hermitean operator. 
Born's rule is not invoked: We make no claim that $p_k$'s are probabilities, or even relate them to amplitudes. We only recognize that -- in addition to pure states given by state vectors -- there are 
mixed states given Hermitean operators. There is no appeal to the physical significance of reduced density matrices. The whole proof parallels our original pure state case step by step. And -- as before 
-- it leads to the special set of orthogonal stable states, reminiscent of pointer states. It is good to 
derive them without any appeal to probabilities: One can then use them -- without any danger of circularity -- as ``events'' while deriving Born's rule.

On the other hand, we cannot really drop any of the first three postulates. This is obvious for the  superposition principle of postulate (i) and for the unitarity postulate (ii) -- they define quantum theory (although (ii) could be weakened e.g. by allowing antilinear and antiunitary evolutions). One can, however, relax the demands of postulate (iii). It is one of the standard axioms, and, in principle, its demands can be always met. Moreover, the ability to reconfirm outcomes of measurements encapsulates predictability that is essential to introduce the very concept of `a state' to describe ${\cal S}$. So it is appropriate to rely on (iii) in a discussion of quantum foundations. However, its central demand -- that measurements should not perturb the measured system -- is only rarely met in laboratory experiments. It is therefore interesting to consider more general measurement schemes, e.g.,
$$ 
( \alpha \ket v + \beta \ket w )\ket {A_0} \Rightarrow \alpha \ket {\tilde v} \ket {A_v}+ \beta \ket {\tilde w} \ket {A_w} = \ket {\tilde \Phi_{\cal SA}} \ .
\eqno(3)$$
\hocom{sequence of measurements:
$$ \ket v \ket {A_0} \Longrightarrow \ket {\tilde v} \ket {A_v} \  , \eqno(3a)$$
$$ \ket w \ket {A_0} \Longrightarrow \ket {\tilde w} \ket {A_w} \  . \eqno(3b)$$}
When $\ket  {\tilde v}$ and $\ket {\tilde w}$ are related with their progenitors 
by a transformation that preserves scalar product (e.g., by a dynamical evolution in a closed system)
proof of orthogonality goes through unimpeded. Both unitary and antiunitary transformations 
are in this class. We can also imagine situations when this is not the case -- $\bra v \ket w \neq \bra {\tilde v} \ket {\tilde w}$. Extreme example of this arises when the state of the measured system retains no memory of what it was beforehand (e.g. $ \ket v \Rightarrow \ket 0, \ \ket w \Rightarrow \ket 0$). Then the apparatus can (and, indeed, by unitarity, has to!) ``inherit'' the information that was contained in the state of the system. Clearly, the need for orthogonality of outcomes disappears. Of course, such measurements do not fulfill axiom (iii) -- they are not repeatable. An example of this situation is
encountered in quantum optics. Photons are usually absorbed by detectors, and coherent 
states (which are not orthogonal) are the outcome states.

It is also interesting to consider sequences of information transfers;
\begin{eqnarray}
\ket v \ket {A_0} \ket {B_0} \dots \ket {E_0} \Longrightarrow
\ket {\tilde v} \ket { A_v} \ket {B_0} \dots \ket {E_0} \Longrightarrow \dots \nonumber \\
\dots \Longrightarrow \ket {\tilde v} \ket {\tilde A_v} \ket {\tilde B_v} \dots \ket {E_v} \ , \nonumber \ \ \ \ \ \ (4a)
\end{eqnarray}
\begin{eqnarray}
\ket w \ket {A_0} \ket {B_0} \dots \ket {E_0} \Longrightarrow 
\ket {\tilde w} \ket { A_w} \ket {B_0} \dots \ket {E_0} \Longrightarrow \dots \nonumber \\
\dots \Longrightarrow \ket {\tilde w} \ket {\tilde A_w} \ket {\tilde B_w} \dots \ket {E_w} \ . \nonumber \ \ \  \ \ (4b)
\end{eqnarray}
Such ``von Neumann chains''  [10] appear in quantum measurements, environment-induced decoherence, amplification, and in quantum Darwinism. 
As information is passed along this chain, links can be perturbed
(as indicated by {\it ``tilde''}). Unitarity implies that -- at each stage -- products of overlaps must 
be the same,
$$ \bra v \ket w = \bra {\tilde v} \ket {\tilde w} \bra {\tilde A_v} \ket {\tilde A_w} \bra {\tilde B_v} \ket {\tilde B_w} \dots \bra {E_v} \ket {E_w} \ . \eqno(5)$$
When a logarithm of both side is taken;
\begin{eqnarray}
 \ln |\bra v \ket w|^2 & = & \ln |\bra {\tilde v} \ket {\tilde w}|^2 + \ln| \bra {\tilde A_v} \ket {\tilde A_w}|^2 + \nonumber \\
& + & \ln | \bra {\tilde B_v} \ket {\tilde B_v}|^2 + \dots + \ln | \bra {E_v} \ket {E_v}|^2 \ . \ \ \  \nonumber (6)
\end{eqnarray}
Therefore, when $\bra v \ket w \neq 0$, as the information about the outcome is passed along 
the two von Neumann chains, the quality of the records must suffer: The sum of logarithms above 
must equal $\ln |\bra v \ket w|^2$, and the overlap of the two states is a measure of their distinguishability.
For orthogonal states there is no need for such deterioration of the quality of records; $\ln |\bra v \ket w|^2 =- \infty$, so arbitrarily many orthogonal records can be made. 

\section{Discussion}

Selection of a preferred basis implied by postulates (i)-(iii) brings to mind einselection -- choice of preferred pointer states [15]. Amplification is precisely such a multiplication of records.  Quantum 
Darwinism [5,~16,~17] is based on the observation that decoherence typically leads to redundant records. 
The main difference 
is the point 
of departure: Our results above follow from very basic assumptions, and lead to very general conclusions. 
Moreover, we only establish that {\it some} orthonormal basis is needed to define measurement outcomes, while einselection shows what specific pointer basis will emerge given e.g., the coupling 
with the environment.

Pointer states einselected for their resilience (in accord with the original definition [15], or its generalization in terms of predictability sieve [5,~6,~19]) are {\em not} guaranteed to 
diagonalize the reduced density matrix  of the system, $\rho_{\cal S}=Tr_{\cal E} \ket {\Psi_{\cal SE}} \bra {\Psi_{\cal SE}}$, where $\ket {\Psi_{\cal SE}}= \alpha \ket v \ket {\varepsilon_v}+ \beta \ket w \ket {\varepsilon_w}$ by analogy with Eqs. (1) and (3). This is easy to see: States that are left unperturbed by the information transfer will coincide with the Schmidt states of 
$\ket {\Psi_{\cal SE}}$ (that are on the diagonal of  $\rho_{\cal S}$) only when their records in the  environment are perfect -- only when scalar products $\bra {\varepsilon_v} \ket {\varepsilon_w}$ of the corresponding `record states' 
vanish.

We take this to mean that only in that case can one safely attribute the usual probability interpretation to 
events associated with pointer states~[22,~23]. This can be illustrated in a setting where the system 
${\cal S}$ is first entangled with the apparatus and then $\cA$ is decohered by its environment. 
Orthogonality inferred from the resilience of pointer states in spite of the immersion of ${\cal A}$
in ${\cal E}$ is then imposed on the pointer states of the apparatus (so that they can retain measurement records). Decoherence will suppress off-diagonal terms representing quantum coherences between records stored in pointer states [15]. This allows for the description of e.g. ${\cal AS}$ correlations in terms of classical probabilities, as is indicated by the vanishing of {\it quantum discord} [18] -- of the difference between two classically equivalent definitions of mutual information that arises in quantum setting.

Resilience of pointer states is especially important in the `environment as a witness' paradigm [5,~16,~17] which recognizes the indirect manner used by the real world observers 
(such as a reader of this text) to acquire information. Observers do not perform direct 
measurements -- reader is not {\em directly} interacting with this page. Rather, he or she acquires information by intercepting small fraction of the photon environment that was emitted by 
(or scattered from) the computer screen (or print on a sheet of paper). Thus, von Neumann's model of direct measurement [10] involving quantum system and isolated quantum apparatus is a gross oversimplification. Even inclusion of the environment to account for 
decoherence (clearly a step in the right direction) does not capture the essential role of 
${\cal E}$: Environment is not just a ``garbage disposal'' for the unwanted quantum coherence, but a communication channel and a witness to the state of the system. Observers acquire their information about the universe by intercepting fragments of $\cE$.

Quantum Darwinism -- proliferation of selected information about the system throughout 
the environment -- is a frequent byproduct of decoherence. Decoherence is caused by monitoring of 
the system by its environment. Quantum Darwinism leads to deposition of redundant records -- multiple ``copies'' of the selected states of the system. And only states that survive this monitoring intact can lead to such {\it redundant} imprinting in the environment.
Existence of multiple records accounts for objective nature of pointer observables, and explains (through arguments put forward elsewhere [5,~16,~17]) why superpositions of pointer states are inaccessible.
Exploring these connections promises to be fruitful, and is especially relevant in the context of interpretations of quantum mechanics that do not invoke explicit ``collapse'' [20,21,22]. 

In presence of einselection reduced density matrix is {\it locally} equivalent to 
probability distribution over events defined by pointer states: Local observers cannot tell when such 
a mixture represents a state that is already known to someone else (and in that sense definite), or 
when it is a part of a larger entangled whole (and, hence, unknown in principle to anyone [22,23]).
This  ``quantum principle of local equivalence''  can be compared to the famous equivalence of the gravitational pull and {\it bona fide} acceleration: When it is locally impossible to distinguish two 
globally distinct situations (e.g., speeding up in an rocket, and being at rest near a mass), such 
local symmetry implies equivalence of all local physical consequences. 

There is a different setting in which symmetric assumptions lead to asymmetric outcomes: 
In symmetry breaking phase transitions, a symmetric initial state and a symmetry 
- preserving equations lead to a configuration that breaks that symmetry.  In 
thermodynamic transitions this choice can be attributed to thermal fluctuations. In quantum phase transitions spontaneous symmetry breaking is sometimes blamed on 
``quantum fluctuations", but this is a meaningless statement. Classical (e.g., thermal) fluctuations 
can break symmetry of the problem, so in an ensemble of thermodynamic systems {\it average} over 
all possible fluctuations can be symmetric, but each member of can be a subject to an ``asymmetric'' fluctuation. By contrast,  ``quantum noise" is a symmetric {\it superposition} of all possibilities, 
so choice cannot be explained any more than  selection of a specific 
outcome can be explained by postulates (i) - (iii). 

In quantum phase transitions -- e.g., in quantum Ising model (Ref. [24]), where decrease of external 
bias leads from a unique ground state $\ket {\rightarrow \rightarrow \rightarrow \rightarrow \rightarrow \rightarrow \dots}$ aligned with a bias field to two broken symmetry  ``ferromagnetic'' options 
$\ket{\uparrow \uparrow \uparrow \uparrow \uparrow \uparrow \uparrow \uparrow \dots}$ and $\ket{\downarrow \downarrow \downarrow \downarrow\downarrow \downarrow\downarrow \downarrow \dots} $ -- the true ground state is degenerate. Any superposition of these alternatives will do for an
infinite system, although in a finite case symmetric $\ket{\uparrow \uparrow \uparrow \uparrow \uparrow \uparrow \uparrow \uparrow \dots} + \ket {\downarrow \downarrow \downarrow \downarrow\downarrow \downarrow\downarrow \downarrow \dots} $ is energetically favored.
Our analysis of symmetry breaking -- the essence of the collapse postulate (iv) -- fits a similar 
framework. Postulates (i)-(iii), when used to express information transfer, imply symmetry breaking: 
They show that many copies can be made, but only of a certain (orthogonal) subset of all possible 
states. These alternatives provide a `menu' -- a set of choices -- but (as a consequence 
of linearity implied by postulate (ii)) quantum theory stops short of selecting any one of them.

Observer who is local -- who can access only a part of the whole system -- will only be able 
to distinguish between the broken symmetry vacua: Starting from any $\ket \phi = \alpha \ket{\uparrow \uparrow \uparrow \uparrow \uparrow \uparrow \uparrow \uparrow \dots} + \beta \ket {\downarrow \downarrow \downarrow \downarrow\downarrow \downarrow\downarrow \downarrow \dots} $, measurements that access less than all of the spins will only be able to correlate 
his records to $\ket{\uparrow \uparrow \uparrow \uparrow \uparrow \uparrow \uparrow \uparrow \dots}$
or $ \ket {\downarrow \downarrow \downarrow \downarrow\downarrow \downarrow\downarrow \downarrow \dots} $,
one of the two broken symmetry states.
The global state can be detected only by a global observer -- someone who forgoes all local measurements, and makes the right measurement of the whole -- measurement of an observable with 
eigenstate $\ket \phi$. Other measurements don't commute with $\kb {\phi }{\phi}$. Therefore, they
``reprepare'' the global state.

Similar conclusions hold in the less ordered but qualitatively similar case of {\it branching states}~[17],
e.g. $\ket \varphi = \alpha \ket v \ket {\varepsilon^{(1)}_v} \ket {\varepsilon^{(2)}_v} \dots \ket {\varepsilon^{(n)}_v} + \beta \ket w \ket {\varepsilon^{(1)}_w} \ket {\varepsilon^{(2)}_w} \dots \ket {\varepsilon^{(n)}_w}$. Such states naturally arise (e.g., via Eq. (1)) in course of unitary evolutions that lead to decoherence~[5,~16,~17,~22].  
Differences between $\ket \phi$ and $\ket \varphi$ are merely quantitative. For instance, in $\ket \varphi$ 
``copies'' deposited in the subenvironments will be generally imperfect, so it will take more 
measurements to find out the branch. Moreover, information will often spread beyond the primary 
environment -- for instance, information about text printed on a page may reside both in the photon 
environment (in an easy to extract form) and in the scattered air molecules (where it is scrambled 
by scattering). This may affect the ease with which one can extract information about 
the ``system of interest'' from these two environments, but it does not change our basic conclusion:
The essence of the controversial collapse postulate (iv) -- symmetry breaking that makes it impossible to detect an unknown preexisting quantum state -- arises from the uncontroversial 
and generally accepted quantum postulates (principle of superposition, unitarity of evolutions, and 
predictability) in a manner that settles much of the long-standing confusion about the role and origin of ``quantum jumps'' and inevitability of ``collapse'' in quantum measurements.
 
\hocom{Same ``indecision'' is true in quantum phase transitions. There, only broken symmetry states are 
in evidence, and detection of their symmetric superposition is all but impossible when a large number
of subsystems is involved: Only a global measurement on all microscopic subsystems could reveal 
the phase between the two obvious alternatives in the case of quantum Ising. By the same token, 
only global measurements of the whole environment can exhibit coherence between branches 
defined by redundant imprinting of information in quantum Darwinism. We have signaled some of 
the analogies that arise, but investigating them in detail is beyond the scope of this paper. }

I thank Richard Jozsa for discussion that stimulated this paper, and for
comments on its early version.

\smallskip

\noindent{References:}

\smallskip

\noindent [1] N. Bohr, {\it Nature} {\bf 121}, 580 (1928).

\noindent [2] P. A. M. Dirac, {\it The Principles of Quantum Mechanics} (Oxford University Press, 1958).

\noindent [3] W. H. Zurek, {\it Physics Today} {\bf 44}, 36 (1991); see also
an `update', quant-ph/0306072.

\noindent [4] E.~Joos, H.~D.~Zeh, C.~Kiefer, D.~Giulini, J.~Kupsch, and 
I.-O.~Stamatescu, {\it Decoherence and the Appearancs of 
a Classical World in Quantum Theory}, (Springer, Berlin, 2003).

\noindent [5] W. H. Zurek, {\it Rev. Mod. Phys.} {\bf 75}, 715 (2003).

\noindent [6] M. Schlosshauer, {\it Rev. Mod. Phys.} {\bf 76}, 1267 (2004); {\it Decoherence}, 
(Springer, Berlin, 2007).

\noindent [7] E. P. Wigner, pp. 260-314 in Ref. [9].

\noindent [8] E. Farhi, J. Goldstone, and S. Guttmann, {\it Ann. Phys.} (N. Y.) {\bf 24}, 118 (1989).

\noindent [9] J. A. Wheeler and W. H. Zurek, eds., {\it Quantum Theory and
Measurement} (Princeton University Press, 1983).

\noindent [10] J. von Neumann, {\it Mathematical Foundations of Quantum Theory} (Princeton University Press, 1955).

\noindent [11] W. K. Wootters and W. H. Zurek, {\it Nature} {\bf 299}, 802 (1982).

\noindent [12] D. Dieks, {\it Phys. Lett.} {\bf 92A}, 271 (1982).

\noindent [13] H. P. Yuen, {\it Phys. Lett.} {\bf 113A}, 405 (1986).

\noindent [14] C. H. Bennett, G. Brassard, and N. D. Mermin, {\it Phys. Rev. Lett.} {\bf 68}, 557 (1992).

\noindent [15] W. H. Zurek, {\it Phys. Rev.} {\bf D24}, 1516 (1981); {\em ibid.} {\bf 26}, 1862 (1982).

\noindent [16] H. Ollivier, D. Poulin, and W. H. Zurek, {\it Phys. Rev. Lett.} {\bf 93}, 220401 (2004); {\it Phys. Rev.} A{\bf 72}, 042113 (2005).

\noindent [17] R. Blume-Kohout and W. H. Zurek, {\it Phys. Rev.} A{\bf 73}, 062310 (2006).

\noindent [18] H. Ollivier and W. H. Zurek, {\it Phys. Rev. Lett.} {\bf 88}, 017901 (2002).

\noindent [19] J.-P. Paz and W. H. Zurek, in {\it Coherent Atomic Matter Waves, Les Houches Lectures}, R. Kaiser, C. Westbrook, F. David, eds. (Springer, Berlin 2001).

\noindent [20] H. Everett, III, {\it Rev. Mod. Phys.} {\bf 29}, 454 (1957).

\noindent [21] M. Gell-Mann and J. B. Hartle, {\it Phys. Rev.} D{\bf 47}, 3345 (1993); R. B. Griffiths,  
{\it Consistent quantum theory} (Cambridge University Press, 2002).

\noindent [22] W. H. Zurek, {\it Relative States and the Environment: Einselection, Envariance, Quantum Darwinism, and the Existential Interpretation}, arXiv:0707.2832 (2007).

\noindent [23] W. H. Zurek, {\it Phys. Rev. Lett.} {\bf 90}, 120403 (2003); {\it Phys. Rev.} {\bf A 71}, 052105 (2005)



\noindent [24] S. Sachdev, {\it Quantum Phase Transitions}, (Cambridge, 1999).

\end{document}